# A Spatially Resolved Radio Spectral Index Study of the Dwarf Irregular Galaxy NGC 1569

Jonathan Westcott[1],* Elias Brinks[1], Luke Hindson[1], Robert Beswick[2] & Volker Heesen[3].

[1]*Centre for Astrophysics Research, University of Hertfordshire, College Lane, Hatfield, AL10 9AB, UK*
[2]*Jodrell Bank Centre for Astrophysics, School of Physics and Astronomy, The University of Manchester, Manchester, M13 9PL, UK*
[3]*Hamburger Sternwarte, Universität Hamburg, Gojenbergsweg 112, 21029 Hamburg, Germany*

## ABSTRACT

We study the resolved radio-continuum spectral energy distribution of the dwarf irregular galaxy, NGC 1569, on a beam-by-beam basis to isolate and study its spatially resolved radio emission characteristics. Utilizing high quality NRAO Karl G. Jansky Very Large Array (VLA) observations that densely sample the 1–34 GHz frequency range, we adopt a Bayesian fitting procedure, where we use Hα emission that has not been corrected for extinction as a prior, to produce maps of how the separated thermal emission, non-thermal emission and non-thermal spectral index vary across NGC 1569's main disk. We find a higher thermal fraction at 1 GHz than is found in spiral galaxies ($26^{+2}_{-3}$%) and find an average non-thermal spectral index $\alpha = -0.53 \pm 0.02$, suggesting that a young population of cosmic ray electrons is responsible for the observed non-thermal emission. By comparing our recovered map of the thermal radio emission with literature Hα maps, we estimate the total reddening along the line of sight to NGC 1569 to be $E(B - V) = 0.49 \pm 0.05$, which is in good agreement with other literature measurements. Spatial variations in the reddening indicate that a significant portion of the total reddening is due to internal extinction within NGC 1569.

## 1 INTRODUCTION

Radio continuum emission from normal galaxies is made up of two key emission processes that are both closely related to recent star–formation (Condon 1992). Thermal (free–free) radio emission, originating from HII regions that have been ionized by massive stars ($\geq 8 M_\odot$), has been shown to be a direct tracer of instantaneous star–formation that holds across many different galaxy types (Murphy et al. 2012; Tabatabaei et al. 2017). Although thermal radio emission is an ideal, extinction–free tracer of star–formation (Murphy et al. 2011), it is intrinsically faint, limited by its brightness temperature of typically $10^4$ K and is overwhelmed at frequencies $< 30$ GHz by non-thermal (synchrotron) radio emission, and at frequencies $> 200$ GHz by the thermal re–radiation of stellar light by dust. The synchrotron emission originates from relativistic Cosmic Ray electrons (CRe) that have been accelerated in supernova shock–fronts (Krymskii 1977; Axford et al. 1977; Bell 1978a,b; Blandford & Ostriker 1978; Drury 1983; Reynolds 2008) and is much brighter and more accessible at $\sim$ GHz frequencies. Yet its relationship to star–formation is far more complicated than thermal radio emission, as CRe acceleration, diffusion (where the CRe experience losses, mainly due to interactions with magnetic fields in the ISM or Inverse Compton scattering) and escape determine the observed non–thermal radio emission characteristics. This is reflected in the Radio–FIR 'conspiracy' (Bell 2003; Lacki et al. 2010) and the non–linear nature of the SFR – non–thermal radio relation (Condon & Yin 1990; Heesen et al. 2014; Tabatabaei et al. 2017). Despite the complexity of the radio emission from star–forming galaxies, several calibrations for the SFR – radio relation exist in the literature, both empirical

* E-mail: j.westcott3@herts.ac.uk

**Table 1.** Key NGC 1569 properties

| Property | Value | Reference |
|---|---|---|
| $\alpha_{J2000}$ | 04 30 49.0 | – |
| $\delta_{J2000}$ | +64 50 53 | – |
| Galaxy Type | IBm | 1 |
| Distance | 3.4 Mpc | 2 |
| $SFR_{H\alpha}$ (Distance scaled) | 0.59 $M_\odot$ yr$^{-1}$ | 3 |
| Angular Size (Major Axis) | 4′.68 | 4 |
| Angular Size (Minor Axis) | 2′.15 | 4 |
| Position Angle | 118° | 4 |
| Inclination | 63° | 4 |

**Reference List**: 1: de Vaucouleurs et al. (1991), 2: Lelli et al. (2014), 3: Hunter & Elmegreen (2004), 4: Jarrett et al. (2003).

(Bell 2003; Murphy et al. 2011; Heesen et al. 2014; Tabatabaei et al. 2017) and theoretical (Condon 1992; Schober et al. 2017).

Regardless of which emission mechanism is studied, both emission components need to be separated, and their associated uncertainties, before any analysis of them can take place. The ideal separation procedure involves taking radio observations that fully sample radio frequency space, and fitting the thermal and non–thermal components simultaneously. In practice however, only a few (2–3) radio frequencies are available, which results in large degeneracies between the fitted thermal and non–thermal components (Niklas et al. 1997). Furthermore, standard $\chi^2$ fitting procedures may underestimate the true uncertainties associated with the separation (Condon 1992). To counter this problem, authors tend to use extinction–corrected H$\alpha$ emission maps as a proxy for the thermal radio emission (e.g. Heesen et al. 2014), and subtract this contribution to isolate the non–thermal emission. Although this method significantly reduces the degeneracy between the two components, the corrections for galactic foreground and internal absorption can be highly uncertain (Basu et al. 2017), and are not normally properly accounted for in the rest of the analysis. As shown by Tabatabaei et al. (2017), a Bayesian methodology can be used to reliably estimate the uncertainties associated with the separation. Furthermore, this approach also allows prior information to be incorporated to reduce the fit degeneracies (Sharma 2017). This motivated us to develop a Bayesian methodology that uses the H$\alpha$ emission as a prior, to separate the thermal and non–thermal radio emission components in a galaxy on a beam by beam basis.

To test our fitting procedure, we studied the dwarf irregular galaxy NGC 1569 (Figure 1; see Table 1 for a summary of its key properties). NGC 1569 has recently undergone a starburst phase, with analysis of its super star clusters (Hunter et al. 2000) and colour magnitude diagrams (Angeretti et al. 2005; McQuinn et al. 2010) indicating that the starburst phase started $\sim$ 100 Myr ago and finished $\sim$ 10 Myr ago. Observations of the turbulent gas kinematics found in H I observations (Johnson et al. 2012) and evidence for galactic outflows found in H$\alpha$ (Waller 1991), H I (Johnson et al. 2012), radio–continuum (Lisenfeld et al. 2004), and X–ray (Martin et al. 2002) observations provide further evidence for this recent starburst phase. Large-scale H I observations of the sky surrounding NGC 1569 indicate that an interaction with the dwarf irregular galaxy UGCA 92 is the possible cause of the recent starburst (Johnson 2013). Although the starburst phase has finished, the current star–formation rate within NGC 1569 is still high compared to its lifetime average (see Figure 8 in McQuinn et al. 2012), and is higher than that normally found for dwarf galaxies ($\sim$ 0.01 $M_\odot$ yr$^{-1}$; Hunter & Elmegreen 2004; Hunter et al. 2012). These properties make NGC 1569 an ideal candidate to test separation methods for 3 main reasons:

(i) It is nearby, bright and well studied in many wavelength regimes. There is a great deal of ancillary data, which the recovered results can be compared against.

(ii) As most of the recent star–formation occurred in a single burst, there is only one CRe population that is currently producing the majority of the observed non–thermal radio emission. Furthermore, as NGC 1569 is small, few CRe will experience spectral aging before escaping the main disk into the halo (Israel & de Bruyn 1988; Kepley et al. 2010). This is a simpler problem than in a larger spiral galaxy where CRe population mixing and aging make separating the non–thermal component more challenging.

(iii) The current star–formation rate is high, the thermal radio emission is also bright and much easier to detect and isolate.

The paper is structured as follows; in Section 2 we describe our observations, data reduction, imaging method and we discuss the impact that missing flux from short spacing has on our results; in Section 3 we describe our fitting procedure; in Section 4 we describe the results retrieved from our fitting procedure; in Section 5 we estimate the reddening along the line of sight to NGC 1569 and derive equipartition magnetic field strengths. Finally, we present our conclusions in Section 6.

## 2   OBSERVATIONS AND DATA REDUCTION

We observed NGC 1569 with the NRAO[1] VLA at 1.5, 3.0, 5.0 and 9.0 GHz over a 3 yr period (see Hindson et al. inprep for further details). The 33.8 GHz observation was retrieved from the VLA archive (project code 11B-078) and reduced in the same manner as the other datasets (see below). A summary of observations is presented in Table 2.

Each VLA observation follows the same observing strategy: starting with a scan of the primary flux calibrator followed by alternating scans of NGC 1569 and the complex gain calibrator, J0449+6332. The observing runs then finished with an additional scan of the primary flux calibrator. For the majority of observations, 3C138 was used as primary flux calibrator, but for the 5 GHz and 33.8 GHz observations, 3C147 was used instead. In all presented observations, the primary flux calibrator was also used to calibrate the bandpass for each observed spectral window.

Before carrying out any calibration, several flagging procedures were employed to remove as much contaminating radio frequency interference (RFI) as possible. For each dataset, we started with a run of the Common Astronomy Software Applications (CASA, McMullin et al.

| Frequency (GHz) | Project Code | Band | Array Configuration | Bandwidth (MHz) | Observation Date (dd/mm/yyyy) | Time on Source (min) |
|---|---|---|---|---|---|---|
| 1.5 | 14B-259 | L | B | 1024 | 10/02/2015 | 27 |
| 3.0 | 14B-259 | S | B | 2024 | 10/02/2015 | 24 |
| 5.0 | 12A-234 | C | C | 2048 | 16/03/2012 | 35 |
| 9.0 | 14B-259 | X | C | 2024 | 16/10/2014 | 15 |
| 33.8 | 11B-078 | Ka | D | 2048 | 18/12/2011 | 81 |

**Table 2.** Summary of VLA observations.

| Band | Central Frequency (GHz) | RMS Noise ($\mu$ Jy beam$^{-1}$) | Taper Radius (k$\lambda$) |
|---|---|---|---|
| L | 1.26 | 56 | 32 |
| L | 1.78 | 65 | 20.5 |
| S | 2.50 | 71 | 21.5 |
| S | 3.50 | 77 | 29 |
| C | 5.00 | 40 | 19 |
| C | 7.40 | 45 | 21 |
| X | 8.50 | 45 | 31 |
| X | 9.50 | 43 | 33 |
| Ka | 33.80 | 55 | 22 |

**Table 3.** Noise Measurements for the maps used in our analysis. All maps have been convolved to a common resolution of 6.3".

2007) flagger in TFCROP mode to initially remove any obvious RFI in the uncalibrated data. We then carried out an initial phase calibration, selecting channels which were determined to be free of RFI. We used these phase solutions to determine and apply a preliminary bandpass to the data. We then employed the CASA flagger in RFLAG mode on this bandpass–corrected data to remove any additional RFI that the earlier flagging procedure may have missed. Finally, we flagged the first and last 5 channels at the beginning and end of each spectral window to reduce the noise contribution from the edges of the bandpasses. We took the flag tables resulting from this procedure and applied them to the original dataset as a starting point for calibration. This flagging method proved effective at removing the majority of RFI in lower frequency observations (< 5 GHz), where RFI becomes a more prominent issue.

We then calibrated each dataset independently following standard calibration procedures in CASA. After finding and applying initial antenna gain solutions, we then visually inspected the data to manually remove any RFI that had not been flagged in our initial flagging attempts. Each dataset was then phase only self–calibrated to convergence using a model consisting of the clean components obtained by incrementally imaging the data. We then finished with a final round of amplitude and phase self–calibration. In all self–calibration iterations, we placed additional outlier fields on any bright sources that were present within the VLA primary beam.

Once the datasets were calibrated, we checked our flux scale by comparing preliminary maps from our VLA observations with other maps available in the literature. We checked our 1.5, 3.0 and 5.0 GHz calibrations by comparing the integrated flux densities of several unresolved sources that are not related to NGC 1569 with those found in lower resolution maps from the Westerbork Synthesis Radio Telescope (WSRT; published in Kepley et al. 2010). We find that our flux scale is consistent with the literature at these frequencies to within 10%. At higher frequencies however, the smaller primary beam greatly limits the number of sources we can compare. To check the flux scale for these observations, we compare the spectral index ($\alpha$; $S_\nu \propto \nu^\alpha$) derived using all of our available observations to those in the literature for some bright, well studied compact sources within NGC 1569. For example, for frequencies in the 1.5 GHz to 10 GHz range, we derive a spectral index of $\alpha = -0.54 \pm 0.06$ for an unresolved supernova remnant (SNR) located towards the south–east of NGC 1569 (designated NGC1569-38; Chomiuk & Wilcots 2009). This is in excellent agreement with spectral indices derived by Greve et al. (2002) and Chomiuk & Wilcots (2009) who find $\alpha = -0.56 \pm 0.03$ and $\alpha = -0.58 \pm 0.15$ respectively. We do not have any additional checks for the 33.8 GHz fluxscale, however the derived spectral indices for various HII regions found in the H$\alpha$ map presented in Hunter & Elmegreen 2004 have a characteristic flat spectrum ($\alpha \simeq -0.1$), lending confidence that our flux calibration at 33.8 GHz is also correct.

### 2.1 Imaging

All of the presented observations are affected by the lack of short baselines, which is to be expected from the array configurations used in the observations and the apparent size of NGC 1569 (see Table 1). We discuss how this affects the observed spectral indices in Section 2.3.

To ensure that our derived spectral indices are as accurate as possible, it is critical to ensure that each map used in our analysis is sensitive to the same spatial scales. As interferometers sample the Fourier transform of the sky intensity distribution (the $uv$–plane), we can ensure we are sensitive to similar spatial scales by only using samples within a particular region of the $uv$–plane. This range is defined as the inner and outer edges (usually measured in wavelengths) of an annulus, centred on the origin of the $uv$–plane, where samples inside the annulus are used for imaging. In this paper, we limit our analysis to the $uv$–range 3.4 – 30 k$\lambda$ as this range is sampled by all our observations. The drawback to limiting the $uv$–range is that we resolve out a large fraction of the extended emission associated with NGC 1569, greatly limiting our analysis of diffuse emission on scales greater than $\approx 30''$.

As NGC 1569 is bright, we are able to split the observed bandwidth in two equal halves for the 1.5, 3.0, 5.0 and 9.0 GHz observations. By doing this, we almost double our available data points, allowing for a more reliable analysis of NGC 1569's resolved spectral properties. We do not split the 33.8 GHz observations as the thermal emission that this waveband is primarily tracing is faint and the receivers are inherently more noisy than those at lower frequencies. By averaging over the entire 33.8 GHz bandwidth, we maximise our sensitivity in this waveband.

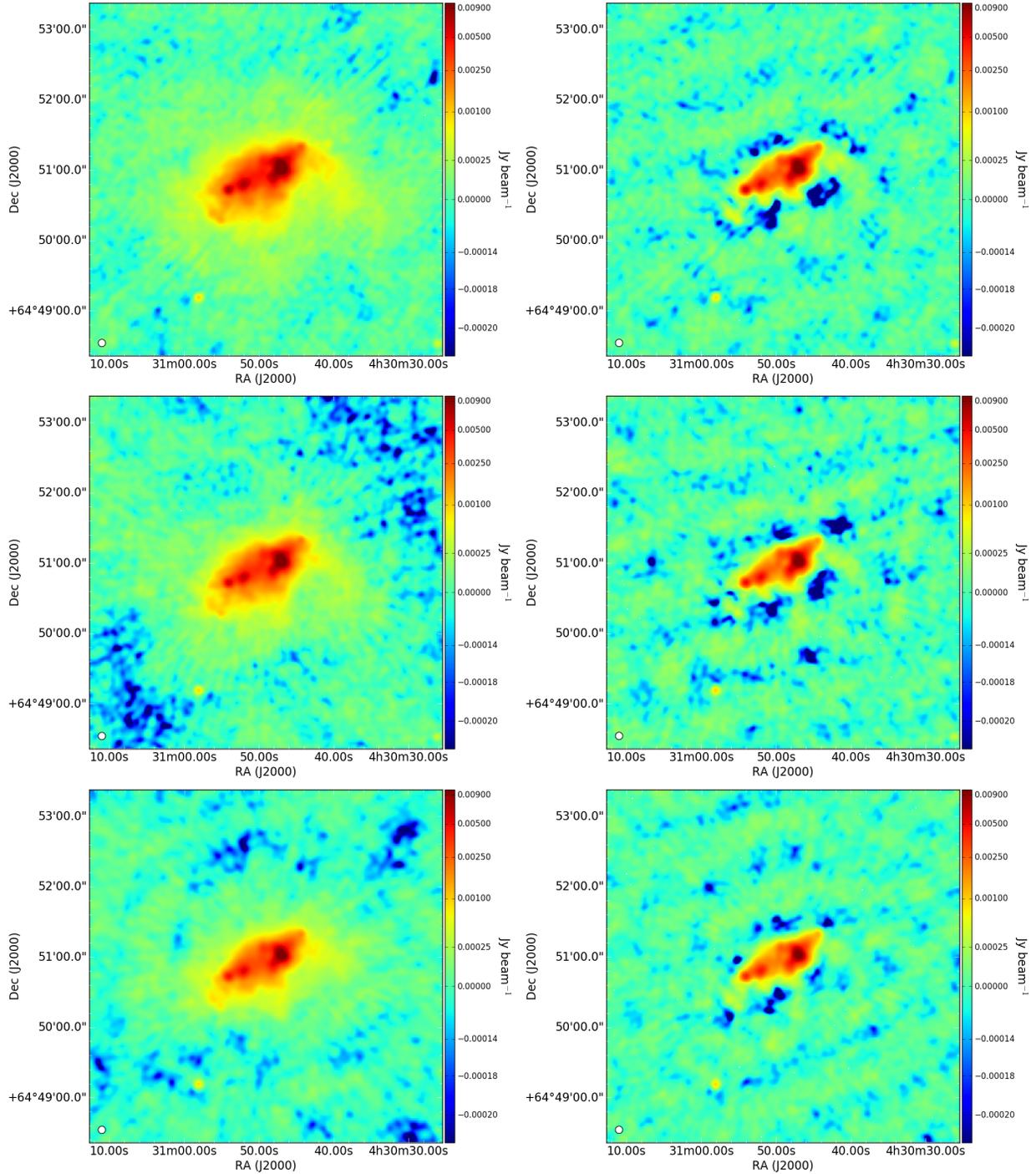

**Figure 1.** Radio maps of NGC 1569 at the frequencies specified in Table 3. The Top maps show NGC 1569 at 1.26 GHz, the Middle maps at 1.78 GHz and the Bottom maps at 2.5 GHz. The maps on the left were imaged using the entire $uv$–range available from the observations in Table 2, and the maps on the right were imaged using the limited $uv$–coverage (3.4–30 kλ). All maps have been corrected for primary beam attenuation and the units for all maps are in Jy beam$^{-1}$. All maps have been stretched to the same log scale to emphasise low surface brightness emission, and all have been convolved to a Gaussian PSF with FWHM 6.3″ (an outline of the beam is shown in the bottom–left of each image).

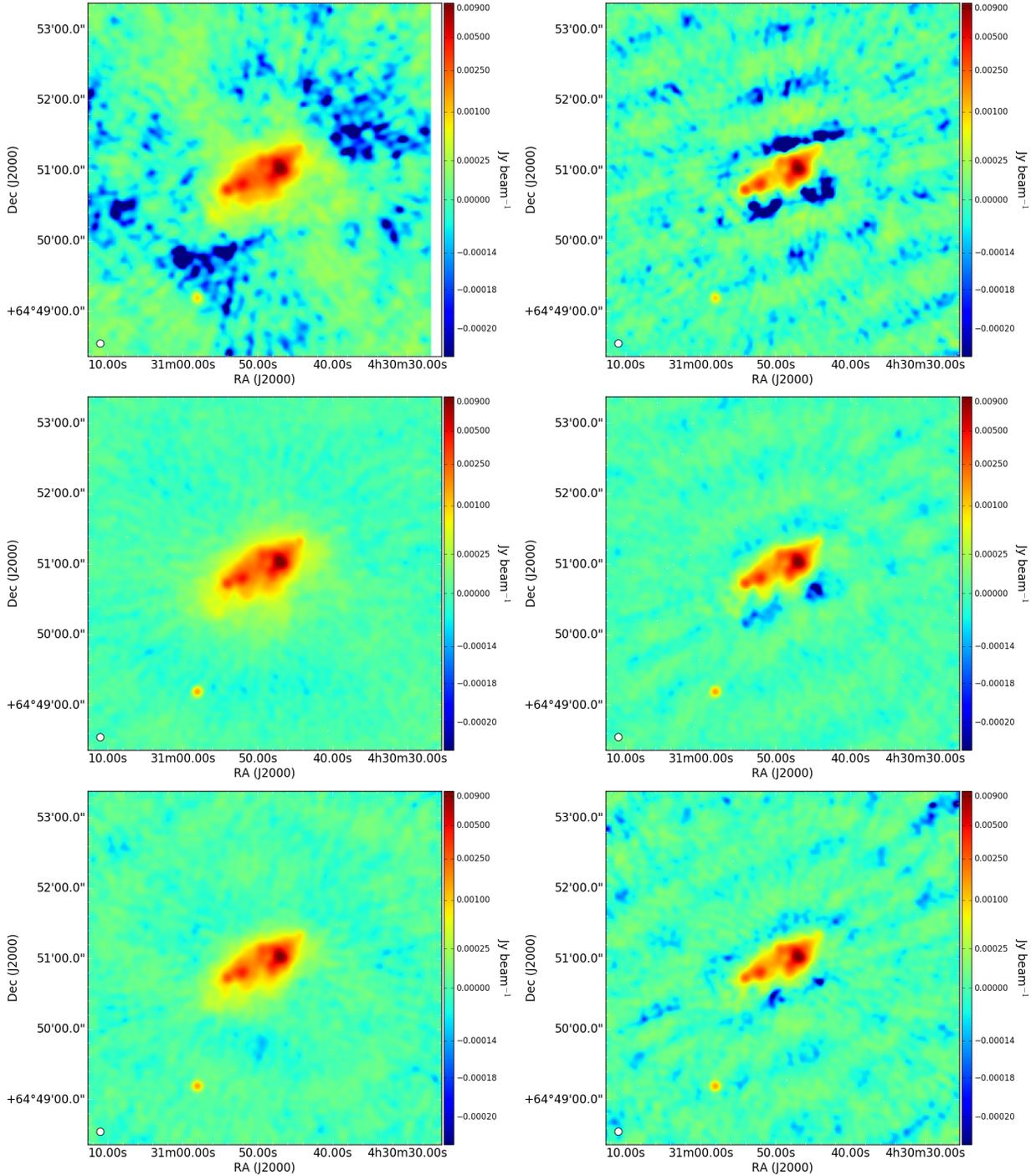

**Figure 1.** Cont. The Top maps show NGC 1569 at 3.5 GHz, the Middle maps at 5.0 GHz and the Bottom maps at 7.4 GHz.

We used multi–scale mutli–frequency cleaning (Rau & Cornwell 2011) to produce our final maps for analysis, 'cleaning' at scales equal to the synthesized beam, twice the synthesized beam and 5 times the synthesized beam with a Briggs robust parameter of 0.5. We additionally apply a circular Gaussian taper whilst imaging to ensure that the final synthesized beams for all maps are as similar as possible. We present the taper parameters in Table 3. Whilst

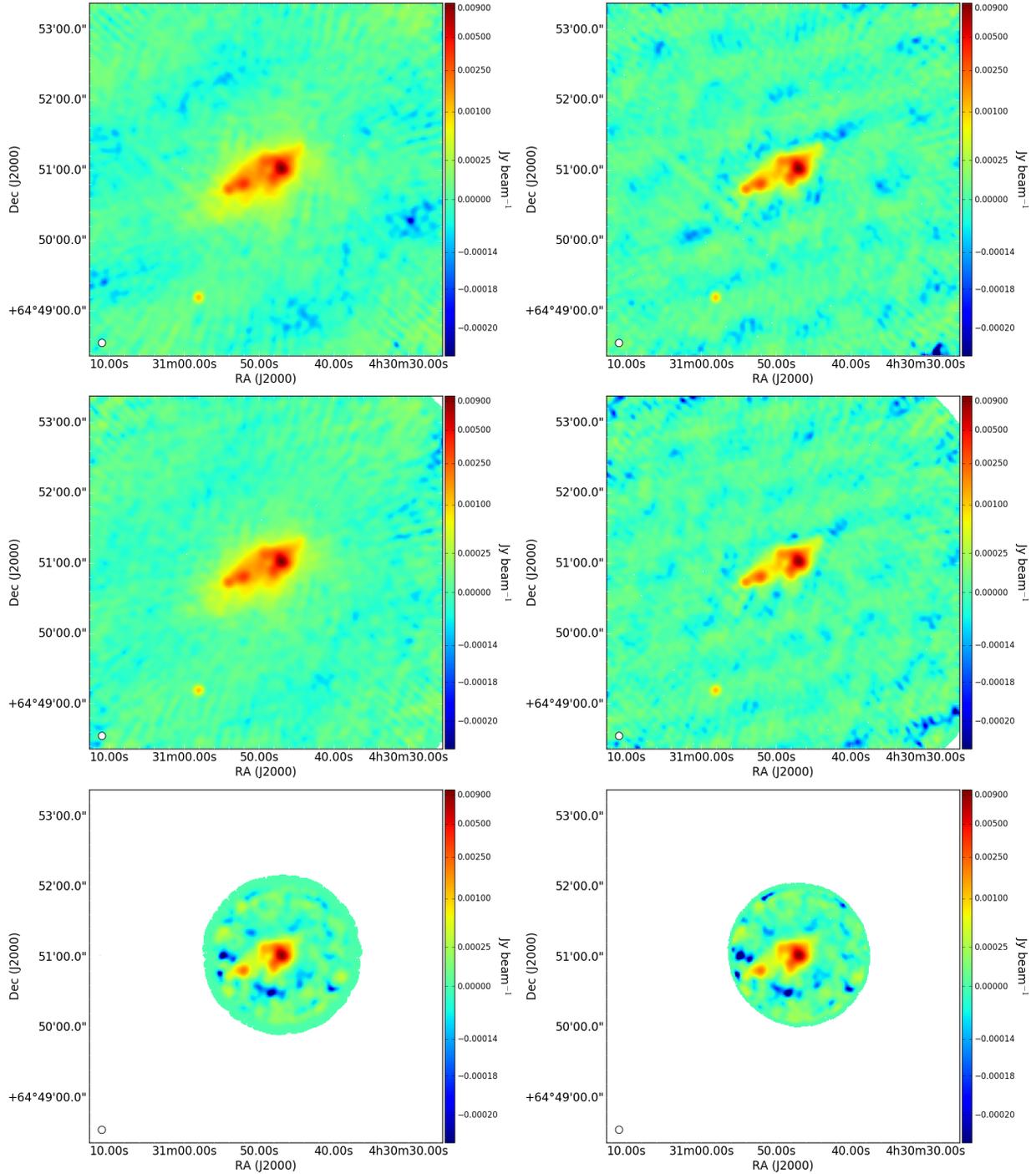

**Figure 1.** Cont. The Top maps show NGC 1569 at 8.5 GHz, the Middle maps at 9.5 GHz and the Bottom maps at 33.8 GHz. We have blanked emission where the primary beam falls to 20% of its peak value. This mainly affects the 33.8 GHz map, where the emission is confined to a circle.

cleaning the 1.5 and 3.0 GHz data, we placed an outlier field on the bright background source, NVSS J042932+645627, to account for its contaminating sidelobes. There was no need to do this for any other background sources as they were sufficiently faint. This was only an issue in these wavebands as this source fell outside the primary beam in the higher frequency observations. Once we imaged all our data, we corrected the maps for the primary beam using the CASA task 'pbcorr' and then convolved all maps to a common resolution. The convolution was carried out using the CASA task IMSMOOTH with the TARGETRES parameter set to True. This uses the clean beam parameters from

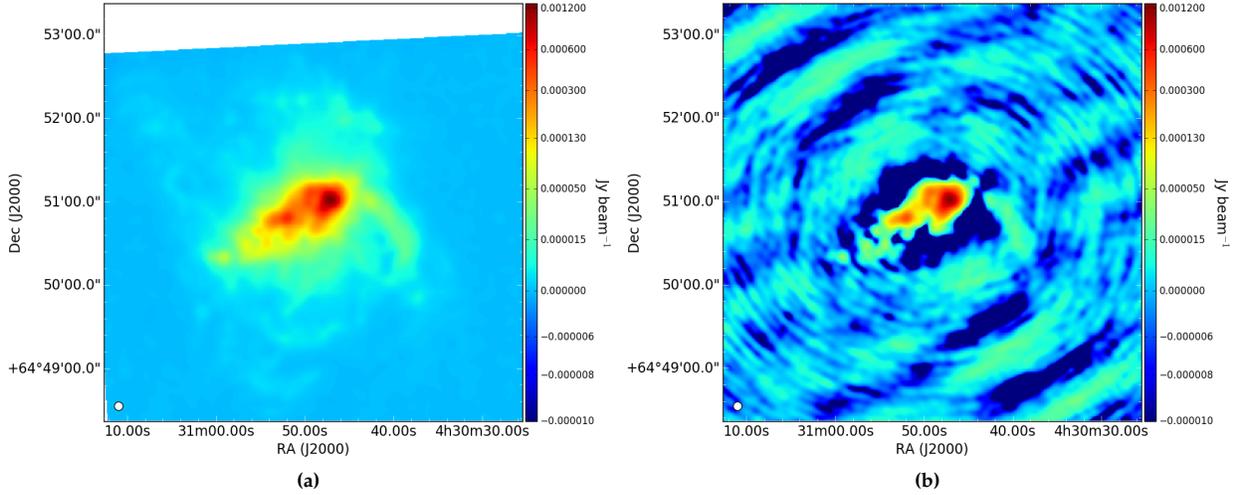

**(a)**                                   **(b)**

**Figure 2.** Models of NGC 1569's thermal emission. Figure 2a presents the thermal emission model derived by directly applying the equations in Deeg et al. (1997) to the Hα map from Hunter & Elmegreen (2004). Figure 2b presents our developed model for the thermal emission, as discussed in Section 2.2. Both maps have been convolved to a Gaussian PSF with FWHM 6.3". Some key differences between these maps include that a large portion of the extended emission from Figure 2a is resolved out in Figure 2b and also the emergence of low–level artefacts from the interferometric imaging procedure.

a given map to determine a convolving kernel that will result in a desired Gaussian point spread function (PSF). We convolved all maps to a circular Gaussian PSF with FWHM 6.3" (corresponding to a linear resolution of 98 pc at 3.4 Mpc). Finally, we regridded all maps to the lowest resolution map. The final radio maps are presented in Figure 1 and the off–source noise levels for each final map are presented in Table 3.

### 2.2 Additional Maps

In addition to the discussed radio continuum maps, we use maps of Hα emission from Hunter & Elmegreen (2004) (see also LITTLETHINGS; Hunter et al. 2012) as a prior on the thermal emission component. We apply the equations from Deeg et al. (1997) to convert from Hα flux to thermal radio continuum flux at 1 GHz. In the conversion, we assume an electron temperature of $10^4$ K (Deeg et al. 1997). It should be noted that this value can vary significantly, as shown in studies of a large sample of galactic H II regions (Hindson et al. 2016). Furthermore, Nicholls et al. (2014) find that an electron temperature of $1.4 \times 10^4$ K is more appropriate in metal poor dwarf galaxies. As the derived thermal emission weakly depends upon electron temperature ($F_{TH} \propto T_e^{-0.34}$), we do not expect that the assumed electron temperature will significantly affect our results. From Hunter & Elmegreen (2004), we assume a 20% systematic uncertainty in the Hα flux scale, which we combine in quadrature with the uncertainty associated with the Hα isolation method (20%, Vilella-Rojo et al. 2015), the uncertainty on the N II contribution (1%) and the photometric uncertainty from the map itself. We do not apply any corrections for galactic foreground or internal extinction at this stage.

As the Hα emission results from the recombination of electrons with protons that have been ionized by Lyman continuum emission originating from massive stars, it is effectively cospatial with the thermal radio emission which originates from the ionized plasma surrounding said massive stars. In other words, the Hα emission effectively has the same spatial distribution as the thermal radio emission. We can therefore create a thermal model that is compatible with the interferometric observations by simulating the corrected Hα map as a radio image with the CASA task 'simobserve'. We 'clean' the simulated dataset with an identical method to that discussed in Section 2.1 and present the developed thermal model in Figure 2. This resulting model is sensitive to the same spatial scales as the interferometric observations, however it still suffers greatly from the effects of extinction. We therefore use the measurements of the thermal emission from these maps as a lower limit in our fitting routine, which we describe in Section 3.3.

### 2.3 The Effects of Missing Flux

Care must be taken when analysing any data from high–resolution interferometric observations as the resulting maps may suffer from problems due to the lack of short baselines in the interferometer array. These problems arise when there is considerable large scale emission that is resolved out by the interferometer, and manifests itself as a negative artefact centred on the source (commonly known as a 'clean bowl'). As radio sources typically have different structures in different frequency regimes (e.g., a halo surrounding a galaxy will appear much brighter and more extended at lower frequency observations than higher frequency ones), it is unlikely that the magnitude and shape of the negative artefacts will be the same at all observed frequencies. Additionally, the interferometer has different responses at different frequencies (e.g., the sampling of the uv–plane will never be identical in any two radio observations at different frequencies), which could further influence measured spectral indices. The combination of these two effects could jeopardize any efforts to measure

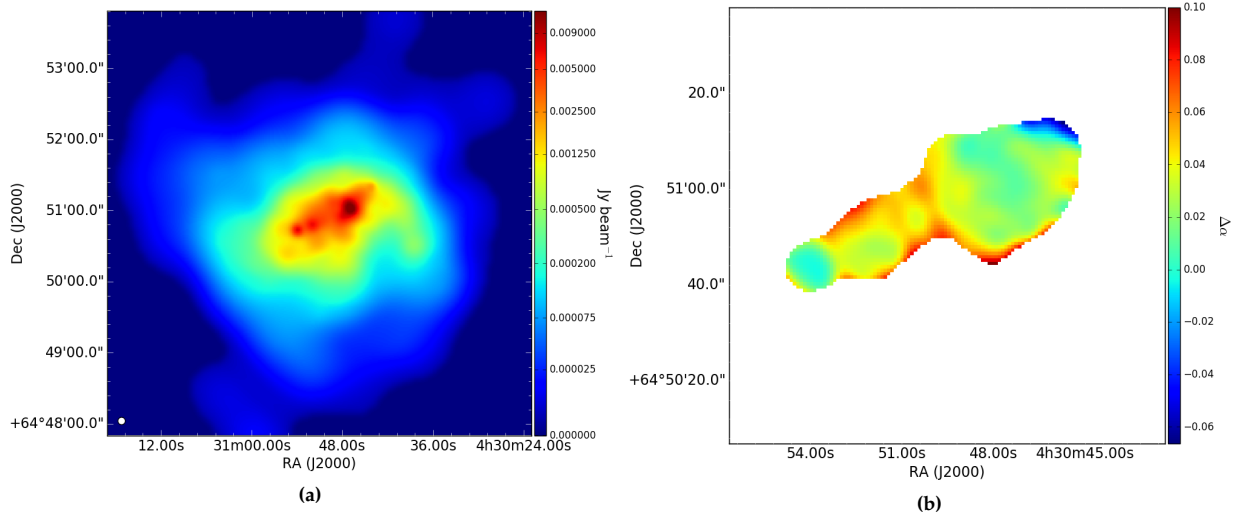

**(a)**                                                                          **(b)**

**Figure 3.** Figure 3a shows the 1.25 GHz model of the radio emission, convolved to a Gaussian with 6.3" FWHM. This model consists of a large scale halo component, a more compact main disk component and some unresolved components within the main disk. Figure 3b presents the residuals after subtracting the total spectral indices recovered from our simulated images from the assumed model. Positive deviations indicate that the total spectral index is flattened whereas negative deviations indicate that the total spectral index is steepening.

spectral indices on a resolved basis. As the presented observations of NGC 1569 do suffer from these effects (see Figure 1), it is necessary to explore how our measured spectral indices may be affected by the interferometric nature of the observations and our chosen imaging procedure.

To test how our measured spectral indices may be affected, we first constructed models of NGC 1569 at the frequencies presented in Table 3. The models consist of 3 distinct components, a large scale halo with a steep spectral index ($\alpha = -0.7$), a more compact main disk component with a flatter spectral index ($\alpha = -0.3$) and a set of high resolution emission components consisting of a collection of unresolved SNRs ($\alpha = -0.5$; Green 2014) and HII regions ($\alpha = -0.1$; Condon 1992). We model NGC 1569's large scale halo by deconvolving the 1.5 GHz WSRT map from Kepley et al. (2010) and isolate the large scale emission using the CASA task 'deconvolve'. To model the main disk component, we image our 1.5 GHz observations with the entire available $uv$–range (1-50 k$\lambda$) and use the clean components that do not correspond to unresolved sources as our disk model. Finally, we take the unresolved sources removed whilst creating the model of the main disk and assign them as either SNR or HII regions as identified in the literature (Chomiuk & Wilcots 2009; Waller 1991). The final maps are constructed by scaling and summing these model components at the frequencies in Table 3. We show the 1.25 GHz model in Figure 3a. Although these single component spectral index models are a simplification of reality, analysing them should identify regions where our spectral index analysis may break down.

We then simulate observations of our model galaxy with the CASA task 'simobserve', using the observing parameters presented in Table 2. We imaged the simulated datasets with the same imaging parameters as detailed in Section 2.1 to produce maps that are comparable to those observed. To measure the resolved spectral indices, we assemble the imaged maps into a datacube. For each pixel in the datacube, we fit a power law by least–squares to measure the resolved spectral index. For comparison purposes, we carry out the same fitting routine for the un–simulated models, which we convolve and regrid to the same resolution.

Figure 3b shows how the recovered spectral indices differ from the input model on a pixel–by–pixel basis. We can see that the spectral index for the bright unresolved sources within NGC 1569 are unchanged and that the spectral index for resolved sections tends to be slightly flatter than the model. The flattening is most pronounced towards the edges of the mask, as this is where the fainter emission begins to compete with the surrounding negative 'clean bowl'. On the other hand, the North–West corner of NGC 1569 appears to be slightly steepened, which is likely an artefact from the 'spur' towards the west of the galaxy. The largest deviation is a flattening of $\Delta\alpha \approx 0.1$, however this only is the case for a couple of pixels. In general it is much smaller with the average deviation being a flattening of order $\Delta\alpha \approx 0.03$. As we carry our fitting procedure on a beam–by–beam basis, we are averaging over the few 'extreme' pixels towards the edge of the mask, which will mitigate their effect on the measured spectral index.

Although in general the difference in spectral index is small, there is considerably more scatter in the 1-10 GHz range. This scatter is due to the interferometer having different responses at different frequencies, primarily due to the different ways the $uv$–plane is sampled.

The flattening tendency can be understood as follows: Although we are resolving out the significant large scale emission from NGC 1569, it is distributed over a large area. On a beam–by–beam basis, this large–scale component contributes little to the detected bright emission from the main disk. As the large–scale halo emission has a steep spectral index, the contribution from the halo would be larger in lower frequency observations than in higher frequencies. When the large–scale emission is resolved out, it corresponds to a larger subtraction of emission in the maps at lower frequencies, resulting in flattening. As the flattening is small across NGC 1569's main disk, it is bright enough such that the missing flux from short spacings does not affect the recovered main disk spectral indices significantly.

This exercise indicates that the systematic errors introduced by missing flux from short baselines and our chosen imaging scheme are small ($\sigma_\alpha \sim 0.03$) when compared with the uncertainties from our fitting procedure (see Figure 4). As we have a wide enough frequency coverage and the parts of NGC 1569 we are spatially sensitive to are bright enough, we can reliably carry out a resolved spectral index analysis on our presented high–resolution observations.

## 3 METHODOLOGY

In this section we discuss the rationale for using a Bayesian methodology to separate the thermal and non–thermal radio emission components, the model that we have chosen to fit and the general fitting procedure. VanderPlas (2014) provides a useful reference on the implementation of Bayesian fitting methods and Hogg et al. (2010) is a useful document describing the merits of using a Bayesian methodology.

### 3.1 Why a Bayesian Method?

In the general case, Bayes' theorem is written as:

$$P(\theta|D) = \frac{P(D|\theta) P(\theta)}{P(D)}, \tag{1}$$

where $\theta$ is a vector describing a set of model parameters and $D$ is a vector representing a set of data points. For the purpose of parameter estimation with a single underlying model, Equation 1 can be simplified to:

$$P(\theta|D) \propto P(D|\theta) P(\theta), \tag{2}$$

where $P(\theta|D)$ represents the posterior probability distribution, $P(D|\theta)$ represents the model likelihood and $P(\theta)$ represents the prior probability distribution.

Equation 2 illustrates two key advantages of using a Bayesian Markov–chain Monte–Carlo (MCMC) fitting procedure over more classical model fitting methods. By using an MCMC approach, a large range of parameter space can be efficiently probed, leading to a statistically robust approximation to the posterior probability distribution. From which, probability distributions for each of the parameters of interest can be retrieved by marginalizing over all other model parameters. Secondly, prior information can be easily included to further constrain fitted parameters.

Both of these benefits should be taken advantage of when attempting to decompose a given radio continuum spectral energy distribution into its thermal and non–thermal emission components. As this type of problem is typically degenerate (Niklas et al. 1997; Callingham et al. 2015; Tabatabaei et al. 2017), classical $\chi^2$ fitting methods may underestimate the true uncertainties associated with the separation (Condon 1992). The marginalized posteriors from the Bayesian approach in this case better reflect the uncertainties associated with the separation process. Furthermore, the addition on prior knowledge can help to reduce the degeneracies between parameters, resulting in better parameter constraints. The main disadvantage of using a Bayesian approach is the required computational resources, but this no longer is a significant hurdle with current technological developments.

### 3.2 Model Definition

As discussed earlier, it is understood that a typical radio continuum spectrum from a purely star forming galaxy is made up of 2 components. A thermal component, originating from ionized plasma surrounding massive stars (HII regions), and a non–thermal component, originating from CRe that are accelerated in SNR shock fronts (see Condon 1992 for a review). As shown by Hindson et al. (2016), HII regions can be assumed to be optically thin at frequencies above 1 GHz. Therefore, an expression for the radio continuum spectrum can be written as:

$$S_\nu = S_T + S_{NT} = A\nu^{-0.1} + B\nu^\alpha, \tag{3}$$

where $S_\nu$ is the measured flux density at frequency $\nu$, $S_T$ is the thermal flux density, $S_{NT}$ is the non–thermal flux density (both in Jy), $A$ is the thermal scaling factor, $B$ is the non–thermal scaling factor, $\nu$ is frequency and $\alpha$ is the non–thermal spectral index. Equations of a similar form to Equation 3 can be derived from physical arguments, assuming that the HII regions are optically thin in the frequency range of interest and that no significant losses have occured (e.g. Murphy 2009; Lacki et al. 2010; Schober et al. 2017). In Equation 3, it is implicitly assumed that there is a single CRe population producing the non–thermal emission. This is a valid assumption to make for NGC 1569 as the majority of the synchrotron emitting CRe would have been accelerated soon after the most recent starburst phase (Angeretti et al. 2005; McQuinn et al. 2010).

Following Tabatabaei et al. (2017), Equation 3 can be parameterised to avoid dependencies of the units of frequency space as:

$$S_\nu = A'\left(\frac{\nu}{\nu_o}\right)^{-0.1} + B'\left(\frac{\nu}{\nu_o}\right)^\alpha, \tag{4}$$

where $A' = A\nu_o^{-0.1}$, $B' = B\nu_o^\alpha$ and $\nu_o$ is a reference frequency. Throughout this paper, we assume a reference frequency of 1 GHz. It then follows that the thermal fraction, $f_T(\nu_o)$, at the reference frequency is given by:

$$f_T(\nu_o) = \frac{S_{T_o}}{S_{\nu_o}} = \frac{A'}{A' + B'}. \tag{5}$$

Although this model is simple, it effectively describes the observed radio emission at mid–radio continuum frequencies (1–10 GHz; Condon 1992; Tabatabaei et al. 2017). There are several additions that could be made to the fitted model. For example, if additional lower frequency observations are included, we could include the effects of free–free or synchrotron self absorption (Tingay & de Kool 2003; Callingham et al. 2015; Hindson et al. 2016; Kapińska et al. 2017). Towards higher frequencies, a break in the non–thermal spectrum due to spectral ageing and inverse Compton scattering (Lisenfeld et al. 2004; Harwood et al. 2013; Heesen et al. 2015) could also be included. Spectral ageing processes are typically seen in the radio halos surrounding normal galaxies (Lisenfeld & Völk 2000) and, as they are unlikely to be significant within the disk of a dwarf galaxy, we do not model it in this current study. Furthermore, a component due to Anomalous Microwave Emission (AME; Draine & Lazarian 1999; Finkbeiner et al. 1999) could be included in the 10-60 GHz frequency range. In the current study, we cannot model this contribution effectively as there is only one data point at 33.8 GHz. We therefore do not model its contribution in this study, but note that if there is significant AME contamination at 33.8 GHz, this will result in an overestimate in the recovered thermal component and our recovered spectral indices will be steeper than the true spectral indices.

### 3.3    Model Fitting Routine

Before we carry out our model fitting routine, we first mask the maps so that we only consider emission that has a significance greater than $5\sigma$ in all observed maps. We then regrid these masked maps onto hexagonal pixels, where each hexagonal pixel has an area that is the same as the synthesised beam. We carry out this procedure because this improves our sensitivity, reduces computational time and also ensures that each hexagonal pixel is more or less independent from each other hexagonal pixel. We determine the uncertainty in the recovered flux density for each hexagon by adding in quadrature a contribution from the noise (presented in Table 3) with an assumed 5% fluxscale uncertainty (Perley & Butler 2017). We finally construct a data–cube out of these hexagonalised maps, with Right ascension on one axis, Declination on the second, and frequency on the third axis.

For each hexagonal pixel in the RA–Dec plane, we slice through the datacube to get a series of flux densities as a function of frequency, to which we fit the model detailed in Section 3.2. Assuming that each map is independent, the flux density uncertainties for each data point are normally distributed and that the uncertainty in the frequency that each map is taken at is negligible, we can use a standard $\chi^2$ objective function to determine the likelihood for a set of measured flux densities given a set of model parameters (Hogg et al. 2010). This probability can be written as follows:

$$P\left(D|\theta\right) = \prod_{\nu} \frac{1}{\sqrt{2\pi\sigma_{\nu}^2}} \exp\left[\frac{-(D_{\nu} - S_{\nu}(\theta))^2}{2\sigma_{\nu}^2}\right], \tag{6}$$

where $D_{\nu}$ is the flux density for a data point in a given pixel at frequency $\nu$, $\sigma_{\nu}$ is the measured uncertainty for that flux density and $S_{\nu}(\theta)$ is the model flux density at frequency $\nu$. $\theta$ corresponds to our given parameters, which are $A'$, $B'$ and $\alpha$ from Equation 4. Equation 6 makes up the first half of the right hand side of Equation 2.

#### 3.3.1    *Assumed Priors*

As with any Bayesian analysis, one needs to be very careful about how priors are defined for the fitted parameters (Gelman 2008). As this step is somewhat subjective, we will describe the functional form of the assumed prior probability distributions for each fitted parameter, and the motivation for these choices in turn.

For the thermal normalisation constant, $A'$, we define an informative prior that depends upon the thermal emission measured from the H$\alpha$ thermal model (see Section 2.2; Figure 2). As the H$\alpha$ emission is vulnerable to extinction effects, we assume that the thermal emission measured from these maps is a lower limit for the true thermal emission. We translate this to a uniform prior with a Gaussian taper at the lower limit:

$$G(A', \mu, \sigma_{H\alpha}) = \begin{cases} \exp\left(\frac{-(A'-\mu)^2}{2\sigma_{H\alpha}^2}\right), & \text{if } A' \leq \mu \\ 1, & \text{otherwise} \end{cases}, \tag{7}$$

where $\mu$ is the thermal emission measured from the thermal model and $\sigma_{H\alpha}$ is the uncertainty on this measurement. This prior is different to a uniform prior that is usually assumed in the literature (e.g. Tabatabaei et al. 2017) as it not only restricts values of $A'$ to positive values, which is physically motivated, but also reduces tails in the posterior probability distribution that stretch to very small $A'$, which improves the constraints on the thermal normalisation constant (see Figure 4).

For the non–thermal normalization constant, $B'$, we adopt a uniform prior that cannot be smaller than 0:

$$H(B') = \begin{cases} 0, & \text{if } B' \leq 0 \\ 1, & \text{otherwise} \end{cases}. \tag{8}$$

This prior is physically motivated as there is no process that can produce a completely absorbed non–thermal spectrum in a dwarf galaxy in the frequency range we are studying. This prior also removes degeneracies where a large thermal normalisation can be coupled with a negative non–thermal normalisation to reproduce the observed spectral energy distribution.

For the non–thermal spectral index, $\alpha$, we note that the non–thermal spectral index for the main disks of star–forming galaxies is typically $\alpha \sim -0.8$ (Condon & Yin 1990) and that the non–thermal spectral index usually varies between $\sim -0.5$ to $\sim -1.2$ due to the injection spectrum of CRe and also due to synchrotron and inverse Compton losses (Longair 1994; Berkhuijsen 1986; Tabatabaei et al. 2017). We therefore define a prior on the non–thermal spectral index corresponding to a broad normal distribution centred on $\alpha = -0.8$ with a standard deviation of 0.4 to encompass this range.

For a given set of model parameters, the prior probability is the multiplication of the values from each prior distribution:

$$P(\theta) \propto G\left(A', \mu, \sigma_{H\alpha}\right) H(B') \exp\left(\frac{-(\alpha+0.8)^2}{0.32}\right). \tag{9}$$

It should be noted that Equations 7 and 8 are not normalised. However, normalising these equations (after assuming an upper limit) will only contribute a multiplicative factor that will be absorbed by the constant of proportionality in Equation 9. As we are only considering a single model, it can be left in its current form.

We use the Python package, EMCEE (Foreman-Mackey et al. 2013), to approximate the posterior probability distributions for each of the fitted parameters. EMCEE makes use of the affine-invariant ensemble sampler for Markov–chain Monte–Carlo presented in Goodman & Weare (2010). We set up 100 'walkers' and allowed each of them to take 1000 steps in parameter space. The initial walker positions were randomly selected from uniform probability distributions. These distributions covered the range between 0 and twice the flux of the lowest frequency data point that is being fit for both $A'$ and $B'$, and between 0 and -2.2 in $\alpha$. We find this was warranted as the lowest frequency data point is normally the brightest and it is very unlikely that the recovered normalizations will be larger than twice its value. We find that the walkers tended to converge after $\sim 100$ steps so we conservatively 'burn in' after 200 steps, yielding 80,000 samples of the sought posterior probability distributions. In the presented results, we quote the 50th percentile of the samples as the best fit, and the 16th and 84th percentiles as the $1\,\sigma$ deviations. An example fit and corresponding corner plot are presented in Figure 4.

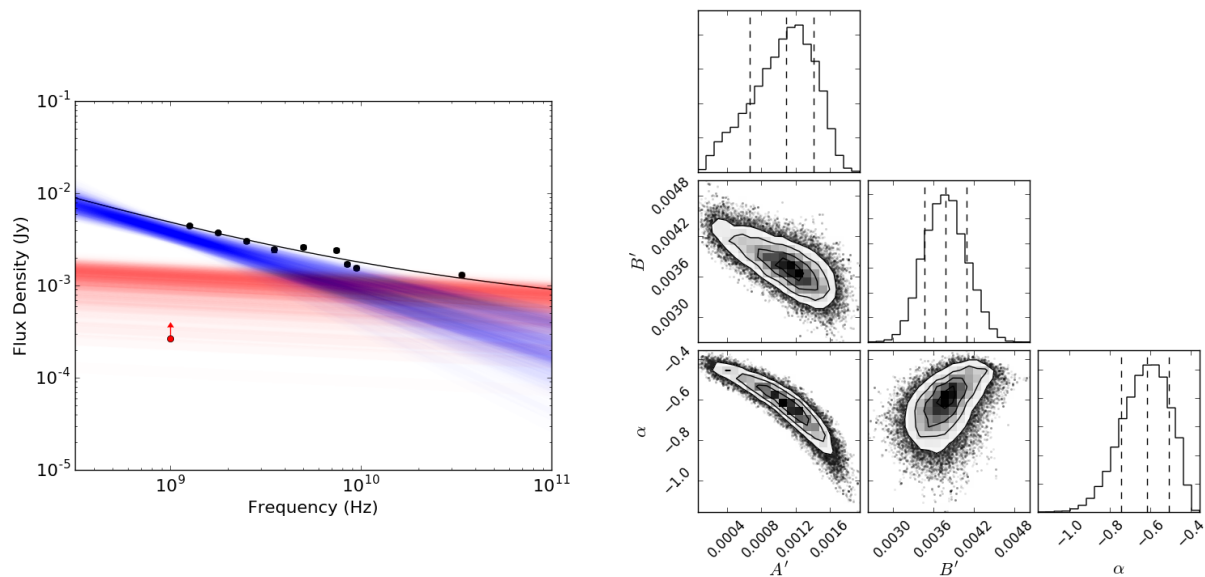

**Figure 4.** An example fit from the procedure outlined in Section 3.3. The left Figure presents an example SED for a hexagonal pixel. The black data points show the measurements from the data–cube; the red point is the lower limit as measured from the thermal model, and the black line represents the best fit to the data. The uncertainties on the flux density measurements are comparable to the size to the black datapoints. We additionally plot 250 models, drawn randomly from the posterior probability distributions, where the red lines represent the thermal component and the blue lines represent the non–thermal component. The Figure on the right presents the corner plot associated with this fit, where the dotted lines correspond to the 16th, 50th and 84th quartiles. For reference, in this fit $A' = 1.09^{+0.32}_{-0.42}$mJy, $B' = 3.78^{+0.32}_{-0.31}$mJy and $\alpha = -0.62^{+0.11}_{-0.13}$.

We test how our assumed priors influence our fits by re–running our fitting procedure, assuming only that the thermal and non–thermal normalisation constants must be larger than 0. In general, the recovered thermal normalisation, $A'$, is slightly larger and the non–thermal normalisation, $B'$, is slightly smaller in the fits assuming minimal priors, with the corresponding uncertainties being similar to the fits with the priors. The recovered spectral indices, $\alpha$, are in general steeper in the fits with minimal assumed priors, with much larger corresponding uncertainties. Within the uncertainties of the fit parameters, both the run with the minimal priors and the run with the full priors return the same results, with the constraints on the recovered spectral index being better with the full priors.

| Frequency (GHz) | VLA Main disk (mJy) | WSRT Main disk (mJy) | WSRT Total (mJy) |
|---|---|---|---|
| 1.26 | 122 ± 6 | 159 ± 8 | 385 ± 20 |
| 2.50 | 91 ± 5 | 124 ± 6 | 269 ± 14 |
| 5.00 | 80 ± 4 | 104 ± 5 | 187 ± 9 |
| 8.50 | 50 ± 3 | 70 ± 4 | 130 ± 7 |

**Table 4.** Comparison of integrated flux density measurements from this study (VLA) and the maps from Kepley et al. (2010) (WSRT). Main disk indicates the integrated flux density over a mask which considers emission brighter than 5 $\sigma$ in the VLA maps (see Figure 1), and Total indicates the flux density measured over the entire galaxy (including the halo). All WSRT measurements were scaled to our observed frequencies assuming a total spectral index $\alpha = -0.4$, and uncertainties were calculated assuming a 5% uncertainty in the flux scale.

## 4 RESULTS

In this section we present the results of the fitting procedure discussed in Section 3.3. In Figure 5, we present maps of the variation of the thermal normalisation, the non–thermal normalisation and the spectral index across NGC 1569's main disk.

### 4.1 Integrated Properties

We obtain NGC 1569's integrated properties by summing the flux density over each of the hexagonal pixels presented in Figure 5. It should be noted that these integrated flux measurements should strictly be treated as lower limits. Due to the interferometric nature of the observations, we are resolving out a large proportion of the emission associated with NGC 1569 on scales larger than ≈ 30″ (see Section 2.3). However, we do recover the majority of the flux from the bright regions within NGC 1569's main disk. This is illustrated if we compare our maps of the main disk (see Figure 1) with Westerbork observations from Kepley et al. (2010), which are sensitive to large scale emission. Integrating over the bright regions that we are sensitive to, we recover ≈ 75 % of the emission at 1.5 GHz, ≈ 70 % of the emission at 3 GHz, ≈ 75 % of the emission at 5 GHz and ≈ 70 % of the emission at 8.5 GHz (see Table 4 for a summary).

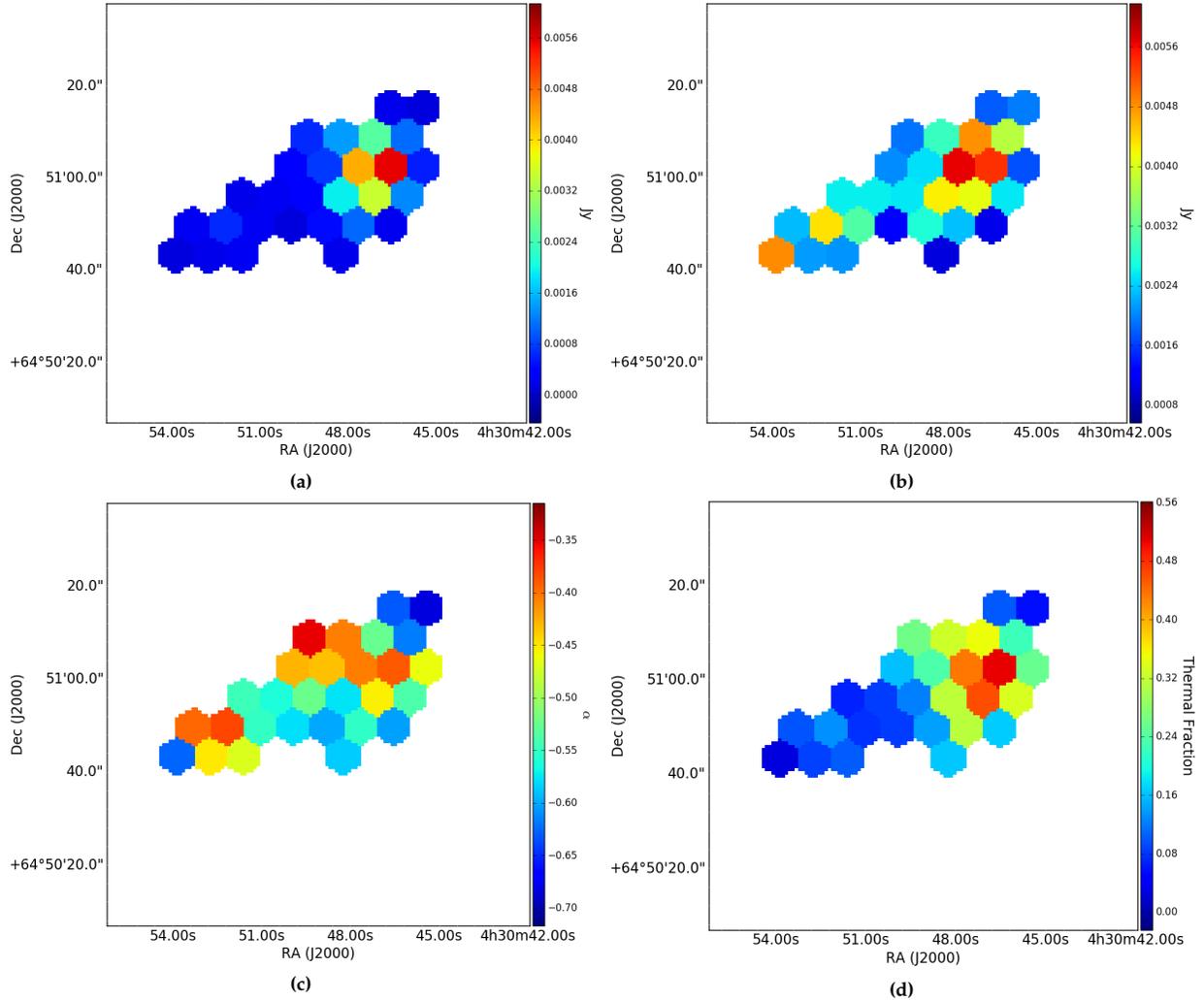

**Figure 5.** Maps of the best fitting model parameters from the fitting procedure discussed in Section 3.3. Each hexagonal pixel has an area that is the same as the synthesized beam. Figure 5a presents the recovered thermal contribution in units of Jy, Figure 5b presents the recovered non–thermal contribution in units of Jy, Figure 5c presents the recovered non–thermal spectral indices and Figure 5d presents a map of how the thermal fraction varies across the disk.

### 4.1.1 Integrated Thermal Emission

As we are typically resolving out 25–30 % of the radio emission from NGC 1569's main disk, it is useful to compare the integrated results from our Bayesian fitting procedure with the literature, after attempting to correct for the missing flux component. If we consider the thermal emission at 1 GHz (see Figure 5a), our Bayesian fitting procedure recovers an integrated flux density of $28.1^{+2.8}_{-3.1}$ mJy. This is significantly larger than the $8.98 \pm 2.7$ mJy we find by integrating the thermal model based on Hα emission over the same area (see Figure 2). This difference indicates that there is significant extinction along the line of sight to NGC 1569, and is discussed further in Section 5.1.

It is difficult to correct this value for missing flux, as we do not know what the relative contributions of the thermal and non–thermal emission are for the missing component. If we assume that the overall thermal fraction within NGC 1569's main disk does not change significantly (see below), this measured value is corrected up to $\sim 50\,\mathrm{mJy}$, which we shall use for comparison with the literature. To first order, we use the H$\alpha$ map from Hunter & Elmegreen (2004), combined with the Spitzer 24$\mu m$ from Bendo et al. (2012) (applying the corrections from Kennicutt et al. 2009) to obtain an independent estimate the total thermal emission over NGC 1569's main disk. Assuming a galactic foreground reddening, $E(B - V) = 0.56$ (Israel 1988), and applying the equations from Deeg et al. (1997), we determine that there is $\approx 56\,\mathrm{mJy}$ thermal radio emission originating from the bright main disk, although this number is sensitive to the assumed reddening from the galactic foreground. Not only is this reasonably close to our corrected thermal emission, it is also in good agreement with the $\sim 55\,\mathrm{mJy}$ that is found by extrapolating the 8 GHz observations of compact H II regions by Israel (1980), but it is significantly smaller than the $\approx 80\,\mathrm{mJy}$ at 1 GHz found from resolved fits to the radio SED by Lisenfeld et al. (2004). This discrepancy could be due to the form of the non–thermal emission model they fit, which includes a break towards higher frequencies, and could recover more thermal emission than the simple power–law model we assume in this study. This exercise shows that our fitting procedure is returning plausible values for the thermal component, indicating that our separation procedure was successful.

For further reference, we obtain an estimate for the upper limit on the total thermal emission from the main disk by assuming that all of the resolved out emission is of thermal origin. This results in a total main disk thermal emission of $\sim 90\,\mathrm{mJy}$. Although this value does not rule the value found by Lisenfeld et al. (2004), it is unlikely that all of the missing flux is of thermal origin as the non–thermal component is typically more diffuse, and therefore more likely to be resolved out than the thermal component.

### 4.1.2 Integrated non–Thermal Emission

Similarly, if we consider the non–thermal emission at 1 GHz, we recover an integrated flux density of $81.0^{+2.9}_{-2.6}\,\mathrm{mJy}$. Again, it is difficult to compare this measurement to other literature values, due to the different spatial scales that the literature radio observations are sensitive to. Lisenfeld et al. (2004) obtain $\approx 80\,\mathrm{mJy}$ for the non–thermal emission integrated across the main disk at 1 GHz, which similar to what we find applying no corrections for missing flux. This is surprising as we would expect the Lisenfeld et al. (2004) value to recover more extended emission than ours, as they have better sensitivity to large scale emission. This further suggests that they may be over predicting the recovered thermal emission within the main disk.

If we again assume that the thermal fraction remains the same for the missing flux, we obtain a corrected non–thermal flux density of $\sim 130\,\mathrm{mJy}$, which we shall use for comparison with the literature. To obtain an independent, first order estimate for the non–thermal radio emission, we scale the disk integrated emission at 1.38 GHz from Kepley et al. (2010) to 1 GHz (see Table 4, assuming an integrated spectral index, $\alpha = -0.4$ (Lisenfeld et al. 2004) to find $\sim 170\,\mathrm{mJy}$ total radio emission at 1 GHz. We then subtract the thermal emission from the main disk found from the combined H$\alpha$ and 24$\mu m$ measurement to isolate the non–thermal emission, resulting in a main disk non–thermal flux density of $\sim 120\,\mathrm{mJy}$. This is again reasonably close to what our Bayesian fitting procedure returns, indicating that our separation procedure is correctly separating the thermal and non–thermal components. Again, for further reference, if we assume that all of the missing flux is of non–thermal origin, we obtain an upper limit for the non–thermal emission of $\sim 140\,\mathrm{mJy}$.

We combine our recovered thermal and non–thermal flux densities to calculate the integrated thermal fraction recovered from the presented observations. Using Equation 5, we obtain a thermal fraction of $0.26^{+0.02}_{-0.03}$ at 1 GHz. This is much larger than is typically found for normal spiral galaxies ($\approx 0.1$; Condon 1992; Tabatabaei et al. 2017), and more closely reflects the higher thermal fractions found from the analysis of other dwarf galaxies (Israel 1980; Klein et al. 1989; Heesen et al. 2011). This conclusion does not change if we calculate the thermal fraction instead using the estimated upper and lower limits for the thermal and non–thermal components. The lower limit for the thermal fraction is $\sim 20\,\%$ and the upper limit is $\sim 50\,\%$, showing that the thermal fraction is certainly higher in NGC 1569's main disk than in normal spiral galaxies. This fraction is likely to decrease however if we include the large scale non–thermal halo surrounding NGC 1569 (Israel & de Bruyn 1988; Kepley et al. 2010). This is because the thermal emission is concentrated in the main disk, whereas the non–thermal emission is spread over a large halo surrounding NGC 1569 (Kepley et al. 2010; see Figure 3a), it is therefore likely that we are resolving out more non–thermal emission than thermal emission in the present observations.

Finally, we measure a disk averaged non–thermal spectral index of $-0.53^{+0.02}_{-0.02}$. This is shallower than typically found in normal spiral galaxies ($\alpha \approx -0.8$; Condon & Yin 1990) and better reflects the spectral index of a galactic SNR ($\alpha \approx -0.5$; Green 2014). This suggests that a young CRe population is responsible for the non–thermal emission that we are detecting within the main disk, and is justified as it is likely that these CRe were accelerated in the most recent starburst phase ($\approx 50\,\mathrm{Myr}$ ago; Angeretti et al. 2005; McQuinn et al. 2010). It should be stressed that this measured non–thermal spectral index is strictly correct only on the spatial scales that the observations are sensitive to. If there is a large–scale component within the main disk that is primarily comprised of an older population of CRe, the measured non–thermal spectral index will be steeper. It is however difficult to estimate the degree of steepening without observations that are sensitive to larger scale emission.

### 4.2 Resolved Properties

NGC 1569's resolved properties are presented on a hexagon by hexagon basis, ensuring that each individual pixel is relatively independent from the others.

The recovered thermal emission (Figure 5a) closely reflects the spatial structure from the prior model (see Figure 2), with a bright, compact source towards the North–West of the galaxy and a fainter source towards the South–East. Both of these sources are spatially coincident with known H II regions detailed in Waller (1991). The morphology of the recovered thermal emission closely reflects that recovered by Lisenfeld et al. (2004), who applied a similar separation procedure but with better sensitivity to larger scale emission. Within the uncertainty of our fits, we recover the same thermal emission in the North–West major H II region as Lisenfeld et al. (2004), as the bright North–West source is compact, compared to the more extended nature of the rest of the disk (see Section 2.3). However, we recover only roughly half of the emission across the rest of NGC 1569's main disk, which could be due to the aforementioned lack of short baselines or the Lisenfeld et al. (2004) fit potentially overestimating the thermal radio emission within the main disk.

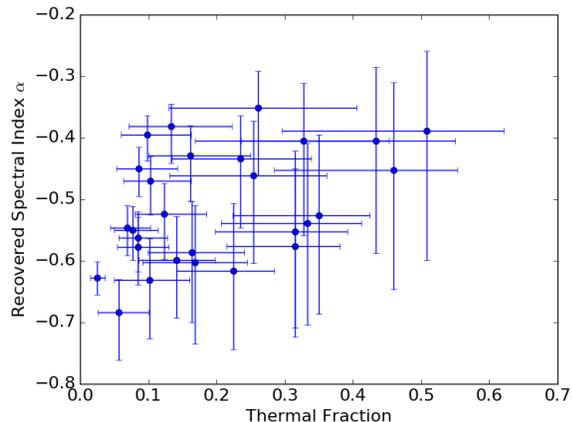

**Figure 6.** Plot of the recovered thermal fraction against the recovered spectral index. Each data point corresponds to a hexagon from Figure 5. The data points are the 50th percentile from the fit posterior probability distributions, and the uncertainties show the 16th and 84th percentiles of the same distribution.

The recovered non–thermal emission (Figure 5b) takes on a more diffuse morphology than the recovered thermal emission. The emission is concentrated in three peaks; One near, yet offset from the major peak in the recovered thermal map, a second coincident with the minor peak in the recovered thermal map and a third towards the South–Eastern tip of the mask. It is not surprising that two of the peaks coincide in both the thermal and non–thermal maps, as this radio emission likely originates from current regions of star formation. It should be noted that the offset peak is coincident with the SNR candidate, N1569-17 (Chomiuk & Wilcots 2009). However, as we are averaging over a relatively large area, it is difficult to separate the compact source properties from the surrounding ISM. The peak towards the South–East is spatially coincident with the well studied SNR, NGC 1569-38 (Chomiuk & Wilcots 2009), and has no corresponding peak in the recovered thermal maps. Here the fitting procedure has correctly identified that this source is dominated by non–thermal emission.

Overall, the non–thermal emission is generally brighter than the thermal emission, which is highlighted in Figure 5d. This Figure shows how the thermal fraction varies across NGC 1569's main disk. The thermal fraction is highest (≈50%) near the current region of major star formation, towards the West of the main disk. The thermal fraction then gradually decreases to ≈15% towards the East of the main disk. The peak in the thermal fraction lines up with the site of on–going major star formation (see the Hα map; Figure 2), which is expected as most compact sources identified within normal galaxies are H II regions, dominated by thermal emission (Condon 1992).

The recovered non–thermal spectral indices are generally between −0.4 and −0.7 across NGC 1569's main disk. As shown in Figure 6, there appears to be a correlation between the recovered thermal fraction and the recovered spectral indices. This is expected as a higher thermal fraction indicates there is recent star formation occurring, and any CRe that are accelerated would be very young. However, as the uncertainties on both the recovered spectral index and thermal fraction are large, we do not detect any significant variation in the recovered spectral index across the main disk.

The variations in the spectral index as seen in 5c are possibly due to two different effects. One is the effect that the missing flux from short spacings has on the recovered spectral indices, which is discussed in more detail in Section 2.3. This likely explains the flattening towards the South–East of the main disk and the steepening towards the North–West. The second effect is that the uncertainties on the recovered spectral indices are much larger in regions with high thermal fractions. This is because it becomes more difficult to constrain the non–thermal properties as the thermal emission becomes more dominant. This likely explains the flattening towards the centre of NGC 1569, although the uncertainties on this measurement are large. As discussed in Section 4.1, the recovered non–thermal spectral indices are consistent with what is expected for a young CRe population.

One additional interesting observation is that the recovered spectral index for the well studied SNR, NGC1569-38, is steeper when the 33.8 GHz observation is included, compared to when it is ignored in the fit. Closer inspection reveals that there is a turnoff in its spectral energy distribution, which may be linked to spectral ageing effects. However, this is beyond the scope of the current paper and will be discussed in more detail in a future paper.

## 5 DISCUSSION

### 5.1 Reddening Estimates

Traditionally, to isolate the non–thermal emission properties of a galaxy, authors generally use scaled Hα emission as a proxy for the thermal emission and, after correcting for extinction effects due to dust, subtract it from the total intensity maps (e.g. see Heesen et al. 2014). There are two main problems that could arise from this procedure; 1) the extinction estimate used to correct for the galactic foreground can be very uncertain, especially in regions at low galactic latitude and 2) applying a single 'blanket' correction for the entire galaxy does not capture any differential extinction that may be occurring within the galaxy itself.

In our fitting procedure, we do not make any assumptions about extinction along the line of sight. Instead we use a thermal model derived from an Hα map that has not been corrected for extinction as a lower limit for the thermal emission component. As we expect our radio observations to be largely free from any extinction effects, by

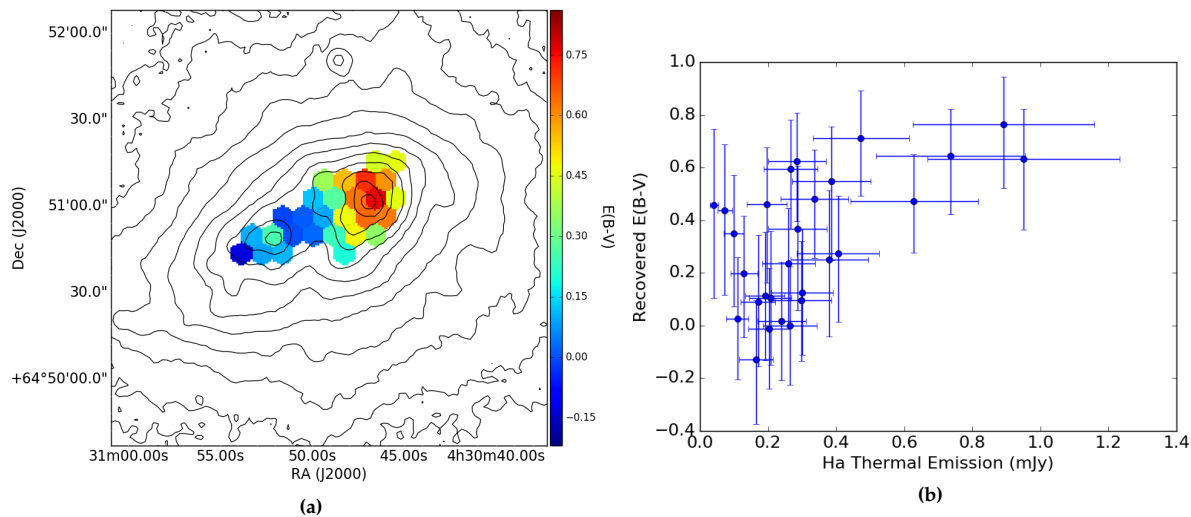

**(a)**                                    **(b)**

**Figure 7.** Figure [7]a shows how the derived reddening, $E(B−V)$, varies across NGC 1569's main disk with overlaid Spitzer $24\mu m$ contours from Bendo et al. (2012). The contours start at 0.36 MJy sr$^{-1}$ and increase in factors of 2. Figure [7]b shows the derived thermal emission measured from the H$\alpha$ map (Figure [2]b) plotted against the recovered reddening.

comparing our recovered thermal emission with the thermal model, we can estimate the reddening along various lines of sight to NGC 1569. From the Cardelli et al. (1989) extinction curves we assume $A_{H\alpha} = 2.54 \times E(B−V)$, which can be combined with Pogsons relation to yield:

$$E(B−V) = \left(\frac{-2.5}{2.54}\right) \log_{10}\left(\frac{S_{H\alpha}}{S_T}\right),\qquad(10)$$

where $S_{H\alpha}$ is the flux density measured from the thermal model at 1 GHz (see Figure [2]b) and $S_T$ is the thermal flux density recovered by our fits at 1 GHz (see Figure [5]a).

The resolved reddening properties are presented in Figure [7]a. There appears to be significant variation across NGC 1569's main disk, with the recovered $E(B−V)$ ranging from as high as $\approx 0.8$ in the main star forming regions to $\approx 0.1$ in some of the more diffuse areas. It should be noted however that the uncertainties on the reddening can be quite large, as shown in Figure [7]b. In Figure [7]a we have plotted Spitzer $24\mu m$ contours from Bendo et al. (2012). The reddening distribution we recover from the radio maps closely follows the warm dust traced by the $24\mu m$ emission, and indicates that the variation we are seeing is due to internal extinction within NGC 1569 itself.

The reddening distribution matches up well with other literature extinction measurements. For the main star forming region (Waller–2; Waller 1991), we find that $E(B−V) = 0.76^{+0.18}_{-0.24}$. This agrees well with literature measurements from the Balmer decrement method, which finds $E(B−V) \approx 0.75$ (Devost et al. 1997; Relaño et al. 2006). Towards the secondary peak in star formation (Waller–7; Waller 1991) we find $E(B−V) = 0.25^{+0.26}_{-0.25}$. This is smaller than that found in the literature, as at a similar resolution to this study, Relaño et al. (2006) find $E(B−V) \approx 0.6$ and a higher resolution study by Devost et al. (1997) finds $E(B−V) \approx 1.0$. Closer inspection of the posterior probability distributions for this tile reveals that the thermal emission is poorly constrained at this location, and that differences in the $uv$–plane sampling are primarily responsible for the significant difference (see Section [4.2]). We also find that the reddening best fit can be negative in tiles where the thermal fraction is low. This primarily is due to the poorer constraints on the thermal emission in these regions of the main disk and are accompanied with larger uncertainties. In order to obtain more precise reddening estimates, longer observations that have been corrected for missing short baselines are required to better constrain the thermal emission component.

We attempt to separate the reddening due to galactic foreground extinction, $E_F(B−V)$, and the reddening due to internal extinction, $E_I(B−V)$, however we stress that the reddening uncertainties are quite large (see Figure [7]b) and hence these results should be regarded as tentative. We estimate the reddening from the galactic foreground by taking the median reddening across NGC 1569's main disk. The motivation for this is that we would expect a uniform reddening correction to exist across the entirety of the main disk for the foreground extinction, and an average over the main disk is biased upwards by significant reddening in the compact main star forming regions. The median recovered reddening is $E(B−V) = 0.33$, which is less than $E(B−V) \approx 0.5$ found by Burstein & Heiles (1982) and Israel (1988). This is possibly due to the Burstein & Heiles (1982) measurement being averaged over a large area of the sky (0.6 deg$^2$), where significant variations on scales of $\sim 100$ pc could occur (e.g. see the dust maps from Schlegel et al. 1998), and the Israel (1988) measurement was made by analysing the integrated properties over the entire galaxy, not taking into account any small scale variations. If we assume the foreground reddening is $E_F(B−V) = 0.3$, we see variations due to internal extinction which can be as large as $E_I(B−V) \approx 0.5$. This reflects the internal extinction corrections that are measured for larger spiral galaxies (Kennicutt 1998; James et al. 2005), suggesting that the star–formation processes may be similar in this region of NGC 1569.

Finally, if we apply Equation [10] to the integrated properties of NGC 1569's main disk to find the average extinction, we obtain $E(B−V) = 0.49 \pm 0.05$. This is in excellent agreement with both Burstein & Heiles (1982) and Israel (1988) who find $E(B-V) \approx 0.5$, although both these studies assume that this is entirely due to the galactic foreground extinction. Figure 7a shows that there must be some internal extinction contributing to the overall reddening, hence these authors may be over–estimating the galactic foreground extinction. As discussed earlier, if we assume the galactic foreground is $E_F(B-V) = 0.3$ we find that the reddening due to internal extinction across the entire galaxy is $E_I(B-V) \approx 0.2$. This closely reflects the internal extinction found for Magellanic irregular and dwarf galaxies (James et al. 2005). These results illustrate the power that radio observations can have in determining resolved extinction corrections that can be used in studies at other wavelengths.

## 5.2 Equipartition Magnetic Field Strengths

The distribution of non–thermal radio emission from a typical galaxy is closely related to its underlying magnetic field strength, as CRe radiate non–thermal radio emission in the presence of magnetic fields (Pacholczyk 1970; Longair 1994; Beck et al. 1996). In principle, magnetic field strengths can be determined from radio observations accompanied with $\gamma$–ray or X–ray observations to gain information on the number density of CRe. However, as these observations are not widely available, an assumption about the relation between CRe and magnetic fields has to be made (Beck & Krause 2005). The most commonly applied approach is to assume that the total energy densities of the CRe and the magnetic fields are approximately equal. This is motivated as the CRe electrons and magnetic fields are strongly coupled, and exchange energy until they reach equilibrium (Beck & Krause 2005). Although equipartition assumptions do not strictly apply to starburst galaxies, Kepley et al. (2010) argue that the magnetic fields derived for NGC 1569 via equipartition assumptions reflect the true magnetic field strengths, based on work by Thompson et al. (2006).

We apply the revised equipartition magnetic field equations derived in Beck & Krause (2005) to obtain resolved magnetic field strength estimates. In the calculation, we assume that the ratio of the CR electron to proton number densities, **K** = 100 (Beck & Krause 2005), that NGC 1569 has a non–thermal emission scale height of 0.4 kpc (Banerjee et al. 2011; Elmegreen & Hunter 2015) and that NGC 1569 is at an inclination of 60° (Jarrett et al. 2003). We further assume that the magnetic fields are predominantly randomly orientated, as no ordered magnetic fields have been observed in deep polarization studies of the main disk (Kepley et al. 2010). Finally, as equipartition magnetic field strength estimates cannot be reliably determined when $\alpha \geq -0.5$ (Beck & Krause 2005), we fix NGC 1569's spectral index to the average over all of the fitted tiles ($\alpha = -0.53$). This is a reasonable assumption to make as, within uncertainties, NGC 1569's spectral index does not vary much from this value within the main disk. Furthermore, this assumption enables a straight–forward comparison of the recovered magnetic field strengths with the literature. We present a map of the recovered equipartition magnetic field strengths in Figure 8.

The spatial distribution of the recovered equipartition magnetic fields is effectively identical to the recovered

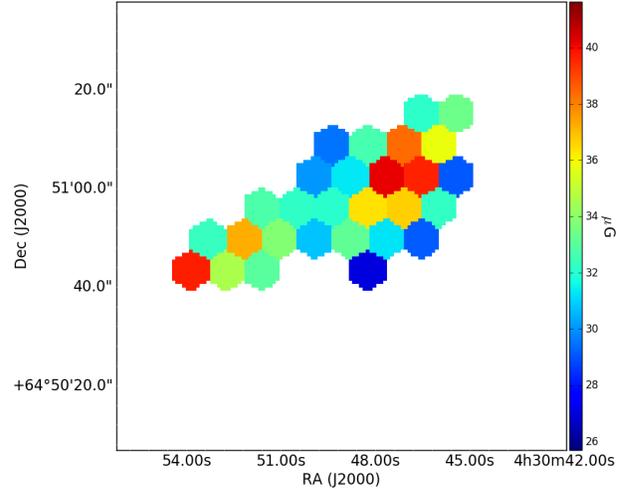

**Figure 8.** Recovered equipartition magnetic field strengths, in $\mu$G, assuming a non–thermal emission scale height of 0.4 kpc. The hexagonal pixels are the same as those in Figure 5.

non–thermal emission, with a peak towards the main star forming region and another cospatial with the SNR NGC1569-38 (Chomiuk & Wilcots 2009). This is because, with a constant non–thermal spectral index, the derived equipartition magnetic field strength is effectively proportional to the recovered non–thermal flux density:

$$B_{eq} \propto \left( \frac{F_\nu}{l} \right)^{\frac{1}{(3-\alpha)}}, \tag{11}$$

where $B_{eq}$ is the equipartition magnetic field strength, $F_\nu$ is the recovered non–thermal flux density at frequency $\nu$, $\alpha$ is the non–thermal spectral index and $l$ is the path length through the non–thermal emitting medium.

The recovered equipartition magnetic field strengths are largest in the region of major star formation and in the SNR NGC1569-38, where it reaches $37.9^{+2.8}_{-2.3}\mu$G and $37.4 \pm 0.4\,\mu$G respectively. The magnetic fields are weakest in more diffuse regions of the main disk, with the lowest magnetic field strength measured being $25.5^{+0.6}_{-0.7}\mu$G. Averaged over the entire main disk, we find an equipartition magnetic field strength of $32.1 \pm 0.3\,\mu$G. We emphasize that the quoted uncertainties are derived from the uncertainties on the non–thermal normalisation only, so the true uncertainties are likely to be much larger. Furthermore, as with most conclusions drawn from high–resolution interferometric observations, these magnetic field strengths should formally be regarded as lower limits due to missing flux. However, as the recovered magnetic field strength weakly depends upon the recovered flux (see Equation 11), we expect this to have a small effect on the recovered magnetic fields.

The recovered magnetic field strengths are in good agreement with those found by Kepley et al. (2010). In the peak associated with the main star forming region, Kepley et al. (2010) find a magnetic field strength of $38\mu$G compared to the $37.9^{+2.8}_{-2.3}\mu$G found in this study. However, these measurements are much larger than those found in other post–starburst dwarf galaxies, although this is probably related to the assumed disk scale height. Assuming a disk scale height of 1 kpc for example will reduce the recovered magnetic field strengths by 33 % (see Equation 11). Chyży et al. (2016) assume a disk scale height of 1 kpc whilst studying the post starburst dwarf irregular IC 10. They find that the peak magnetic field strengths in regions of on–going star–formation are as high as 29 μG. We obtain a very similar magnetic field strength if we assume the same scale height ($29.2^{+2.2}_{-1.6}$ μG), although it should be noted that Basu et al. (2017) find from their SED analysis that these regions in IC 10 can be modelled as being dominated by thermal emission, meaning the uncertainty in the Chyży et al. (2016) result could be large. Regardless, the recovered magnetic field strengths are much larger than that found in other dwarf galaxies (Chyży et al. 2011; Roychowdhury & Chengalur 2012) and in larger Magellanic–type galaxies (Jurusik et al. 2014). This is possibly due to either the large number of SNR occurring after the starburst, generating and amplifying the magnetic fields within NGC 1569 via the small scale dynamo (Chyży et al. 2011, 2016), or because of the fact that this resolved study is focused on high surface brightness emission whereas the older literature studies are mainly based on emission integrated over the entire galaxy, or a combination of the two.

Although our recovered magnetic field strengths likely underestimate the true magnetic field strength due to missing flux, it is unlikely that it is significantly affecting our results as the equipartition magnetic field strengths vary weakly with recovered flux (see Equation 11). If we are missing 30 % of the flux due to missing short spacings, our recovered equipartition fields will only underestimate the true equipartition magnetic field strengths by ∼ 10 %. We find that the magnetic field strengths are similar to those found in turbulence dominated spiral arms and central regions in normal galaxies, where the magnetic field strength is ≈ 20 − 30 μG (Basu & Roy 2013; Beck 2016). This suggests that the relativistic environment in NGC 1569's main disk could be similar to that found in a spiral arm in a normal star–forming galaxy.

# 6 CONCLUSIONS

We apply a Bayesian methodology, where we use maps of the Hα emission that have not been corrected for extinction as a prior in the fit, to high resolution VLA maps of the dwarf irregular galaxy, NGC 1569, in order to separate its thermal and non–thermal radio emission components on a resolved basis. Our main conclusions are as follows:

(i) On an integrated basis, we recover $28.1^{+2.8}_{-3.1}$ mJy thermal radio emission and $81.0^{+2.9}_{-2.6}$ mJy non–thermal radio emission at 1 GHz, although these values should be treated as lower limits. We estimate that we are resolving out ≈ 30 % of the main disk flux density at all observed frequencies, hence the relative spectral properties (thermal fraction and non–thermal spectral index) are preserved. These measurements correspond to a high thermal fraction of $0.26^{+0.02}_{-0.03}$ at 1 GHz, which reflects what is found in other studies of dwarf galaxies. However, if the entire galaxy is considered, this fraction should be treated as an upper limit. It is likely to decrease slightly if the halo, which is dominated by non–thermal emission, is included.

(ii) On a resolved basis, we find that the recovered thermal emission closely follows the structure of the prior Hα map

and is consistent with the separation carried out by Lisenfeld et al. (2004). There are two main peaks in the thermal distribution, which both match up with known HII regions. We recover less thermal emission in the Western peak than previous measurements, which is likely due to differences in the sampling of the *uv*–plane between observations. The recovered non–thermal emission is more diffuse than the recovered thermal emission and is also generally brighter. The non–thermal emission consists of three emission peaks, two of which line up with known regions of current star formation and one corresponding with the known SNR NGC 1569-38. The resolved thermal fraction is ≈ 15% across most of the main disk, and increases to ≈ 50% in the region of major star formation.

(iii) The resolved spectral indices are generally between α = −0.4 and −0.7 and tend to be flatter in regions of active star formation. There appears to be a correlation between the recovered spectral index and the recovered thermal fraction. However, within uncertainties, the resolved non–thermal spectral index does not vary significantly across NGC 1569's main disk. The average non–thermal spectral index across the main disk is α = −0.53 ± 0.02. This spectral index is shallower than that found for normal spiral galaxies, and closely reflects what is found for young SNR. This indicates that the CRe population producing the non–thermal emission is young and probably originates from the recent starburst phase the galaxy has undergone.

(iv) By comparing the prior Hα map that has not been corrected for extinction with our recovered thermal emission map, we estimate the reddening along the line of sight to NGC 1569. Integrated over the main disk we find that the overall reddening is E(B − V) = 0.49 ± 0.05. This is in excellent agreement with the galactic foreground estimates from the literature. However, variations in the reddening across NGC 1569's main disk indicate that internal extinction is significantly contributing to this estimate. By taking the median reddening across the main disk, we estimate that the reddening due to the galactic foreground is $E_F(B−V) ≈ 0.3$, with internal extinction contributing on average $E_I(B−V) ≈$ 0.2. On a resolved basis however, the variations in internal extinction can be as large as $E_I(B−V) ≈ 0.5$, which is similar to that found in larger spiral galaxies.

(v) Using the recovered non–thermal emission characteristics, we derive estimates for the equipartition magnetic field strengths. Assuming a scale height of 0.4 kpc, we find that our recovered equipartition magnetic field strengths vary between 25 μG to 38 μG across the main disk. Our recovered magnetic field strengths are similar to those found in the post–starburst dwarf irregular galaxy, IC 10, but are in general larger than those found for dwarf galaxies. These magnetic field strengths are similar to those found in the spiral arms and central regions of normal spiral galaxies, indicating that the relativistic ISM may be similar in both.

Future studies applying the presented separation procedure should ideally use maps that have been corrected with single dish observations. Not only would this improve the constraints on the fitted parameters by reducing the scatter in the galaxy SED, it would also simplify the interpretation of the recovered fit parameters.

## ACKNOWLEDGEMENTS

JW acknowledges support from the UK's Science and Technology Facilities Council [grant number ST/M503514/1].

EB and LH acknowledges support from the UK Science and Technology Facilities Council [grant number ST/M001008/1].

We thank Ute Lisenfeld for providing us with a map of the thermal radio emission from NGC 1569, taken with the the Cambridge Ryle Telescope at 15.4 GHz, which was used to confirm the robustness of our approach to separate thermal and non-thermal emission. We thank Martin Hardcastle, Donna Rodgers–Lee and Martin Krause for careful reading of the manuscript and useful comments. We also thank the anonymous referee for their useful comments and suggestions.

This paper has been typeset from a TeX/LaTeX file prepared by the author.